\documentclass[11pt]{article}
\usepackage{graphicx}
\usepackage{setspace}

\begin{document}

\title{Assessing the spatial and field dependence
of the critical current density in 
YBCO bulk superconductors by scanning Hall probes}
\author{F. Hengstberger\thanks{E-mail: \texttt{hengstb@ati.ac.at}},
M. Eisterer, M. Zehetmayer and H. W. Weber\\
\footnotesize Atomic~Institute~of~the~Austrian~Universities,
Vienna~University~of~Technology,\\
\footnotesize Stadionallee~2,
1020~Vienna,
Austria}
\date{}

\date{}
\maketitle
\begin{abstract}
Although the flux density map of a bulk superconductor
provides in principle sufficient information for calculating 
the magnitude and the direction of the supercurrent flow,  
the inversion of the Biot-Savart law is ill conditioned for thick samples,
thus rendering this method unsuitable 
for state of the art bulk superconductors.
If a thin ($< 1\,\mathrm{mm}$) slab is cut from the bulk,
the inversion is reasonably well conditioned
and the variation of the critical current density in the sample
can be calculated with adequate spatial resolution.
Therefore a novel procedure is employed,
which exploits the symmetry of the problem
and solves the equations non-iteratively,
assuming a planar $z$-independent current density.
The calculated current density at a certain position
is found to depend on the magnetic induction.
In this way the average field dependence of the critical current density $J_c(B)$ 
is obtained also at low fields,
which is not accessible to magnetisation measurements
due to the self-field of the sample.
It is further shown that an evaluation of magnetisation loops,
taking the self-field into account,
results in a similar dependence in the field range accessible to this experiment.

\end{abstract}
\vfill

\section{Introduction}
At present, 
bulk superconductors with several centimeters in diameter
and about one centimeter in thickness,
trapping remanent magnetic fields exceeding \mbox{1\,T} at \mbox{77\,K},
can be reproducibly grown~\cite{car06}.
Texturing of the monolith is achieved by top seeded melt texture growth (TSMG), 
where crystallisation evolves from a seed crystal placed on top of the bulk.
Five growth sectors are formed, 
propagating from the facets of the crystal 
through the entire material~\cite{car98}.
The critical currents achieved in each growth sector,
especially as a function of the seed distance,
are therefore of particular interest.

A straightforward approach is to cut the sample 
and to characterise small pieces by magnetometry.
However,
this procedure is destructive
and takes extensive measurement time.
On the other hand,
scanning Hall probe techniques can be employed
to analyse the local properties of the bulk~\cite{hai05}.
Among them the magnetoscan technique~\cite{eis03} proved to provide 
detailed information on the critical current flow on a local scale.
However, 
even the strongest permanent magnets used in the magnetoscan device
activate currents only in the uppermost layer of the bulk.
The information refers to depths of less than~\mbox{1\,mm}~\cite{zeh06}
and the inside of the bulk cannot be probed. 

Currents flowing in the entire volume of the bulk 
can be activated by performing a (zero) field cooled hysteresis loop in a magnet.
In principle sufficient data to calculate the current density
on the sub-mm length scale, 
where substantial changes in $J_c$ are expected,
can be obtained from trapped field maps.

Although elaborate and numerically stable techniques exist~\cite{wij96,joo98,car03}, 
it was shown \cite{eis05}
that due to the large thickness of those samples 
the matrix inverted in such procedures is notoriously ill conditioned
and can even be numerically singular.
Thus,
a reduction in thickness,
either by grinding or preferably by cutting the bulk into disks, 
is mandatory to assess the local critical current distribution
in this way.
If the disk is sufficiently thin, 
a scanning mesh can be found
which allows both
an analysis of the current density on the sub-mm scale
and a comparison to the magnetoscan signal at certain positions.

\section{Numerics}
\subsection{General}

\begin{figure}
  \centering
  \includegraphics[width=.4\textwidth]{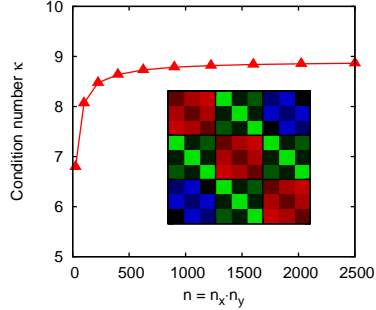}
  \caption{\label{fig:cond}
    ({\it Color Online})
    Estimate of the condition number for the parameters used in the experiment
    (\mbox{$s=0.2\,\mathrm{mm}$}, \mbox{$d=0.55\,\mathrm{mm}$}).
    Since the matrix exceeds \texttt{Scilab} memory stack,
    the condition number was calculated
    for increasingly larger systems of up to $50 \times 50$ points.
    The convergence of the condition number is evident.
    {\it (Inset)} Symmetry of the matrix involved in the computation.
    The nine Toeplitz blocks constitute a block Toeplitz matrix.}
\end{figure}

\begin{figure}
  \centering
  \includegraphics[width=.4\textwidth]{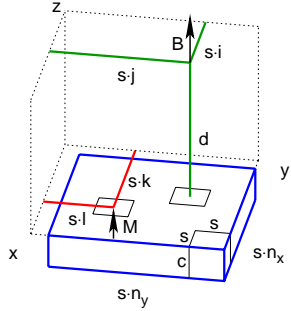}
  \caption{\label{fig:csystem}
    Coordinate system employed.}
\end{figure}

In the coordinate system used in the following
(cf. figure~\ref{fig:csystem})
the top sample surface lies in the $x,y$-plane
and the perpendicular component of the magnetic induction 
$B=\vec{B}\cdot\vec{e}_z$ is assessed.
Similar to~\cite{wij96} the current density is expressed as 
the curl of a magnetisation density pointing in $z$-direction
$\vec{J}=\vec{\nabla}\times M\,\vec{e}_z$.
This implicitly satisfies current conservation $\vec{\nabla}\cdot\vec{J}=0$
for the $z$-independent planar current distribution $\vec{J}(x,y)$.
Unlike in~\cite{wij96}
the basic equation of the problem is derived
by splitting the magnetic induction into two components
and using a scalar magnetic potential
(see the Appendix for a detailed derivation).
Discretisation of the integral equation
using cubic volume elements with constant $M$ 
results in the 2D matrix equation

\begin{eqnarray} 
  \sum_{k=1}^{n_x}\sum_{l=1}^{n_y} K_{i,j,k,l}M_{k,l} & = & B_{i,j} \label{eqn:onea} \\
  K_{i,j,k,l} & = & F(\Delta x,\Delta y,\Delta z)\; 
  \Big|^{s(i-k+\frac{1}{2})}_{s(i-k-\frac{1}{2})}\; 
  \Big|^{s(j-l+\frac{1}{2})}_{s(j-l-\frac{1}{2})}\;
  \Big|^{d}_{d+c} \label{eqn:oneb} \\
  F(\Delta x,\Delta y,\Delta z) & = &  
  \frac{\mu_0}{4\pi}\tan^{-1}\,(\frac{\Delta x \Delta y}
       {\Delta z\sqrt{\Delta x^2+\Delta y^2+\Delta z^2}})\,.
       \label{eqn:onec} 
\end{eqnarray}

Here $c$~denotes the sample thickness,
$d$~the distance between the active area 
of the Hall probe and the top sample surface (gap), 
$s$~the step width,
and \mbox{$i,k=1\ldots n_x$}, \mbox{$j,l=1\ldots n_y$}
the indices on the mesh;
the antiderivative $F$ is evaluated
at the eight corners of the cubes in~(\ref{eqn:oneb}).

Equation~\ref{eqn:onea} can be mapped one-to-one to 1D
by substituting 
$i'=i+j\,(n_x-1),\ k'=k+l\,(n_y-1)$.
The equation now reads

\begin{equation}
  \sum_{k'=1}^{n_x n_y}K_{i',k'}'M_{k'}=B_{i'}
  \label{eqn:two}
\end{equation}

and the problem can be tackled by matrix inversion algorithms.

As pointed out earlier,
the matrix $K_{i,j,k,l}$ is in fact a
Toeplitz block Toeplitz matrix~\cite{wij96}. 
This is a consequence of the translation invariance of the Biot-Savart law 
($K_{i,j,k,l}=K_{|i-k|,0,|j-l|,0}$)
and results in a highly symmetric matrix $K_{i',k'}'$
(cf. inset of figure~\ref{fig:cond}).
Therefore, 
both efficient storage and a fast algorithm for solving the system can be expected.
However,
\cite{wij96}~exploits the symmetry only for the storage of the matrix elements 
and the method of conjugated gradients using the fast Fourier transform (FFT) 
is employed to solve the linear equation.
Although the procedure is fast, 
employing FFT \cite{wij96,joo98}
implies the unnecessary assumption of periodicity in $B$
outside the measurement area,
which may create artefacts,
if currents are flowing close to the edge of the scanning area.

Block Toeplitz matrices occur in a number of problems,
such as image reconstruction or system identification. 
Fortunately,
an efficient and fast algorithm,
which exploits the symmetry of the structured matrix,
is provided in~\cite{nic08}.
The computation time is approximately 10~minutes
for a $200 \times 200$ system on a desktop PC.
This is presumably longer than FFT based algorithms, 
but still much less than the actual measurement time. 

\subsection{Condition number}
\label{sec:condition}
The residuum of the solution is of the order of the machine error.
However,
it is shown in standard numerical algebra textbooks,
that matrix inversions can amplify 
a relative (measurement) error $\epsilon_B$
in the right-hand side of~(\ref{eqn:two}),
leading to an unknown error in the calculated magnetisation density $\epsilon_M$.
This behaviour is described by the condition number $\kappa$ of a linear system

\begin{equation}
  \|\epsilon_M\|_2\leq\kappa(K')\cdot\|\epsilon_B\|_2\,,
  \label{eqn:three}
\end{equation}

where the errors are measured in Euclidean norm.
It was shown in~\cite{eis05}
that the condition number of the inverted matrix is in notoriously high 
for bulk superconductors.
Therefore,
special attention has to be paid to the choice of the step width~$s$,
defining $\kappa(K')$ at a given gap~$d$.
With a realistic distance~$d$ of about \mbox{0.15\,mm} 
(see section~\ref{sec:exp}) 
and a sample thickness of \mbox{0.55\,mm},
the condition number was estimated
using the numerical computation package \texttt{Scilab}~\cite{sci08}
(cf. Fig.~\ref{fig:cond}).

As a rule of thumb one aims at a condition number $\kappa$ such that

\begin{equation}
\|\epsilon_M\|_2 < 1
\label{eqn:four}
\end{equation}

holds.
It was found that the matrix inversion is reasonably well conditioned
as long as the step width is larger than the gap.
The actual choice of \mbox{$s=0.2$\,mm} is adequate to
resolve the spatial variations of the critical current on a sub-mm scale.
The condition number $\kappa \approx 9$
combined with a rough estimate of the relative measurement error 
\mbox{$\|\epsilon_B\|_2 \approx 0.01$}
results in $\|\epsilon_M\|_2 \approx 0.09$,
satisfying~(\ref{eqn:four}).

Calculating the current density involves computing a (numerical) derivative,
which implies that the relative error in $\vec{J}(x,y)$
will be high,
if the change in $M(x,y)$ (the current at this position)
vanishes,
as for example outside of the bulk or in defects.
This is a peculiarity of measuring the relative error
of a quantity close to zero.
Note that the absolute error in $\vec{J}(x,y)$ is bounded
by~(\ref{eqn:three})
and expected to be acceptably low
for most of the current distribution inside the sample 
as long as~(\ref{eqn:four}) holds.

\section{Experimental}
\label{sec:exp}
A thin disk was cut from an undoped YBCO bulk superconductor
with a diameter of \mbox{$26.5\,\mathrm{mm}$},
which was grown by the top seeded melt growth technique \cite{kra95},
using a diamond saw.
The cut was made near the upper surface of the bulk
and the disk was polished with abrasive paper
to a thickness of \mbox{$\mathrm{0.55\,mm}$}.

Commercial Hall probes from \textsc{Arepoc}
with active areas of \mbox{$25 \times 25$\,$\mu$m$^2$} (trapped field map) 
and \mbox{$50 \times 50$\,$\mu$m$^2$} (magnetoscan)
were used in the experiments.
All scans were carried out with a step width of \mbox{0.2\,mm},
which allows to apply the inversion (see section~\ref{sec:condition}).

For the trapped field maps a scanning area of 
\mbox{$28 \times 28$\,mm} was used.
The sample was field cooled in a split coil magnet 
in a field of \mbox{1.4\,T}.
In order to minimise relaxation effects during the measurement,
the scan was started \mbox{10\,min} 
after sweeping the magnetic field to zero.
The temperature of the liquid nitrogen bath was recorded 
prior to and after the scans
and was found to be \mbox{77.2\,K}, 
increasing due to oxygen uptake by about \mbox{0.1\,K} during the measurement.
The resistive offset of the Hall probe was determined 
at the end of each measurement,
with the Hall probe still immersed in liquid nitrogen, 
but at a large distance from the bulk.

A somewhat wider scanning area of \mbox{$30 \times 30$\,mm} 
was used for the magnetoscans.
This precaution was taken to assure 
that the magnet would not produce artefacts 
by stopping over the bulk after finishing a single line of the scan.
The magnetoscan was carried out using a small SmCo permanent magnet
of \mbox{2\,mm} in diameter, 
applying an induction of around \mbox{150\,mT}
at the top sample surface of the bulk.
Further details of the technique can be found in \cite{eis03}.

\section{Results}

\begin{figure}
  \includegraphics[width=.4\textwidth]{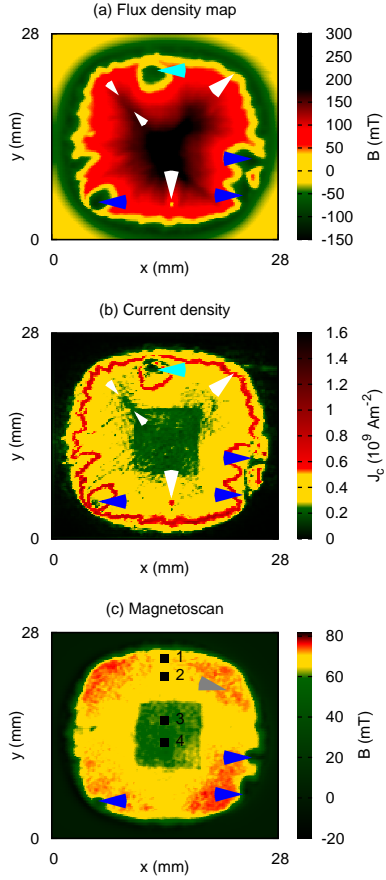}
  \caption{\label{fig:all_comp}
    ({\it Color online})
    {\it (a)}
    A number of extended defects are detected
    in the trapped flux density map (blue arrows)
    and strong negative fields of up to \mbox{$-100\,\mathrm{mT}$} 
    are observed (green area).
    {\it (b)}
    The inversion reveals a strong correlation between the magnetic induction 
    and the current density (white arrows in {\it a},{\it b}).
    In the defects (blue arrows) as well as in the $c$-axis growth sector
    (green rectangular area in the center)
    the critical current density is significantly reduced.
    {\it (c)}
    The $c$-axis growth sector is again detected by the magnetoscan.
    The small length scale of the variations in the signal
    indicates strong local changes in the critical current (gray arrow).
    (The black rectangles indicate the cubes cut for magnetometry.)}
\end{figure}

\subsection{Inversion of the Trapped Flux Density Profile}
Assuming that the variation of the critical current over
the sample thickness can be neglected
immediately implies that the trapped field profiles recorded 
on the bottom and top surface are identical,
which was confirmed by the experiment.
The maxima of the trapped field were equal within one percent
on both sides of the bulk
and found to be \mbox{$252\,\mathrm{mT}$}.
Strong negative fields of up to \mbox{$-100\,\mathrm{mT}$}
were detected close to the edge of the sample,
which result from the high diameter to thickness aspect ratio of the disk
(cf. Fig.~\ref{fig:all_comp}a).

The high reproducibility between several measurements 
shows that the gap between the Hall probe and the sample surface
(\mbox{$\approx 0.15\,\mathrm{mm}$})
remains unchanged in subsequent runs.

The inversion of the trapped flux density profile of the 
bottom surface of the disk is depicted in figure~\ref{fig:all_comp}b.
Most of the bulk carries a current density of around 
\mbox{$4\,\cdot\,10^8\,\mathrm{Am}^{-2}$} in the remanent state.
The defects (blue arrows) close to the edge appear as regions, 
where the critical current is drastically reduced,
most likely due to cracks or large scale inclusions.
A remarkable result of the inversion is the detection
of the $c$-growth sector at the center of the bulk.
It is clearly displayed in the current density map 
as a rectangular area with low critical current density (green region).

In addition,
a clear negative correlation between the critical current density
and the magnitude of the perpendicular magnetic induction $B$ is found,
for example close to one of the $a$--$a$ growth sector boundary,
where both the magnetic field and the current density change simultaneously
(small white arrows).
The correlation is most prominent at low fields,
i.e. high current densities of up to \mbox{$10^9\,\mathrm{Am}^{-2}$}
flow close to the sample edge,
where the magnetic induction changes sign 
and therefore $B \cong 0$.
Moreover, 
a point inside the bulk (lowest white arrow) 
with reduced magnetic induction is reproducibly detected,
where the current density significantly exceeds
the nearby current densities.
This demonstrates that a strong field dependence
of the critical current is present,
especially at low fields.

\subsection{Comparison to Magnetoscan}
Due to the strong field dependence of the critical current,
it is difficult to compare different regions of the bulk,
as the self-field of the bulk is position-dependent
in the remanent state.
Contrary, 
in the magnetoscan the background field of the permanent magnet 
is constant and the self-field is smaller, 
since the currents are activated only 
in an area of about the magnet's diameter~\cite{zeh06}.

All except one of the prominent defects
are found in the magnetoscan (cf.~Fig.~\ref{fig:all_comp}c). 
A possible explanation would be a large defect or inclusion at a depth,
which exceeds the penetration depth of the permanent magnet.
Similar to the inversion,
the reduction of the critical current is evident
from the low magnetic response in the $c$-axis growth sector.
Moreover,
the observable granularity on a sub-mm scale of the bulk's response
indicates strong local variations in the critical current density.

\subsection{Comparison to Magnetometry}

\begin{figure}
  \centering
  \includegraphics[width=.4\textwidth]{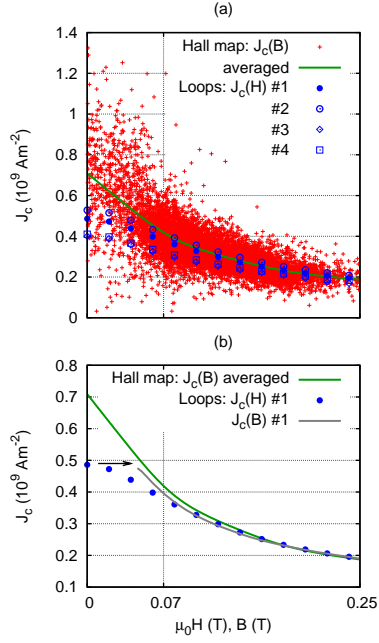}
  \caption{\label{fig:JcB_comp}
    ({\it a}) $J_c(B>0)$ correlation from the inversion,
    showing the average field dependence of the critical current density.
    The magnetisation $J_c(H)$ data is similar to 
    the smoothed data from the current calculation (green line) for high fields.
    The curves differ at low fields,
    because the self-field is not taken into account 
    in the $J_c(H)$ data of the cubes.
    Further, the critical currents vary also from sample to sample
    in the magnetisation data;
    the cubes cut from the seed area show the lowest critical current density.
    ({\it b}) Evaluation of the hysteresis loop for cube \#1
    taking the self-field into account.
    The field generated by the supercurrents in the sample
    shifts the applied magnetic induction to higher fields (arrow),
    resulting in a $J_c(B)$ dependence,
    which is now similar to the average data obtained from the current calculation.
    There is no indication for a flattening of the curve at small fields.}
\end{figure}
 
The calculated current densities can be correlated 
with the $z$-component of the magnetic induction at their position.
For this purpose the magnetic induction in the central plane
inside the bulk was calculated after the inversion (see the Appendix),
resulting in a $J_c(B)$ plot (cf. figure~\ref{fig:JcB_comp}).
Only points with positive induction $B>0$ were considered,
which effectively cuts off the noise
due to the extended defects close to the sample edge.

For comparison four cubes ($2\times2\times0.55\,\mathrm{mm}^3$)
were cut from different positions of the bulk (cf.~Fig.~\ref{fig:all_comp}c)
and analysed in a SQUID magnetometer at the temperature 
of the liquid nitrogen bath during the previous scans.
The magnetisation loops were evaluated assuming the Bean model.
A large sample to sample variation was found, at this point
the lowest critical current density being located 
at the $c$-axis growth sector in the center of the bulk (cubes \#3 and \#4),
in agreement with the above results.
Although numerical errors in the inversion procedure
(mainly due to the numerical derivation)
cannot completely be excluded,
the variation in the magnetisation data elucidates
the scatter in the calculated current densities,
if one takes into account that the averaged volume 
is much larger in the SQUID measurements.
This suggests that the scatter is primarily due to 
strong local variations of the critical current in the bulk,
which is further supported by the granularity observed 
in the magnetoscan signal (cf. Fig.~\ref{fig:all_comp}c).

The average field dependence of the critical current density 
from the magnetisation loops $J_c(H)$
shows good agreement with $J_c(B)$ obtained from the trapped flux density maps,
except at low fields,
where the curves clearly differ (cf. Fig.~\ref{fig:JcB_comp}a).
This can be explained in terms of the sample's self-field,
which is not considered in the evaluation of the magnetisation experiment,
since it takes only the externally applied field $\mu_0H$ into account
and neglects the field generated by the currents flowing in the sample.

To estimate this contribution the magnetoscan signal
serves as a first approximation,
because the currents are induced 
in an area determined by the magnet's diameter
and the volume of current flow mainly contributing to the magnetoscan signal
is therefore similar to the cubes used in magnetometry.
Indeed the average induction of the magnetoscan (\mbox{$\sim70\,\mathrm{mT}$})
is close to the field,
where the $J_c(H)$ and $J_c(B)$ start to differ.

For a more detailed analysis
an evaluation method was applied (again based on the Bean model),
which accounts for both the externally applied field
and the mean self-field in the sample,
thus providing an approximate $J_c(B)$ dependence \cite{wie92}.
Especially for low applied fields,
where self-field effects become dominant,
the evaluation shifts all data points to higher fields
(cf. Fig.~\ref{fig:JcB_comp}b).
This effect is particularly clear for the remanent state
at zero applied field (arrow),
where the magnetic induction is solely due to the trapped self-field.
Consequently,
a sample cannot be probed at zero magnetic induction
by SQUID magnetisation experiments.

When accounting for the self-field contribution,
good agreement between the $J_c(B)$ curve from the SQUID loops 
and the average $J_c(B)$ curve from the inversion
is found in the range accessible to the magnetisation experiment
(depicted in figure~\ref{fig:JcB_comp}b for one of the cubes).
Moreover,
the field,
where the correction starts to become effective
and the self-field becomes important,
is equal to the simple estimate made above using the magnetoscan data 
(\mbox{$70\,\mathrm{mT}$}).

In contrast to SQUID magnetometry,
the current calculation allows one to analyse the $J_c(B)$ dependence
in fields ranging from zero induction to the maximum trapped field.
A clear dependence of the critical current density
on the magnetic induction $B$ is revealed in this way.
There is no indication for a flattening
in the average $J_c(B)$ at low fields. 

\section{Summary}

The inversion of the Biot-Savart law
represents an ill conditioned problem for bulk superconductors,
but parameters,
e.g. the step width of scan and the thickness of the sample,
can be found,
which allow its application to thin disks cut from the sample.
The matrix equation was solved without any additional assumptions
by a fast algorithm,
which exploits the symmetry of the problem.
In this way,
for example the $c$-axis growth sector
was clearly identified as a distinct region of low critical current density,
a result,
which is also obtained by the magnetoscan and
confirmed by magnetisation measurements.

From a technological point of view both methods,
the magnetoscan and the inversion of the flux density map,
are complementary.
The magnetoscan provides information 
on the local critical current density at a certain constant background field,
which is of interest for superconducting bearings.
On the other hand,
information about the remanent current flow at the bulk's self-field
is important for magnet applications
and can be obtained by the inversion of the trapped flux density map
of thin disks cut from the sample.

In addition,
the current calculation was found to provide the important 
average field dependence of the critical current 
also at low fields (below the self-field),
a region,
which is not accessible to magnetic measurements,
even if the self-field is explicitly accounted for
in the evaluation procedure.
There is no indication of a plateau in $J_c(B)$
and the critical current density is found to depend
on the field in a continuous way
also at the lowest magnetic inductions.

\subsection*{Acknowledgement}
The authors wish to thank Vasile Sima for valuable discussions.
This work was supported in part by the Austrian Science Fund
under contract \#17443.

\renewcommand{\theequation}{A-\arabic{equation}}
\setcounter{equation}{0}  
\section*{Appendix}

The magnetostatic Maxwell equations
($\vec{\nabla}\cdot\vec{B}=0$ 
and Ampere's law $\vec{\nabla}\times\vec{B}=\mu_0\vec{J}$\,)
are treated by splitting the magnetic induction into a sum of two fields 

\begin{equation}
  \vec{B}=\mu_0(\vec{\Omega}+\vec{M})\,,
  \label{eqn:aone}
\end{equation}

where $\vec{M}$ is chosen to satisfy the differential equation

\begin{equation}
  \vec{\nabla}\times\vec{M}=\vec{J}
  \label{eqn:atwo}
\end{equation}

in entire space,
which satisfies current conservation $\vec{\nabla}\cdot\vec{J}=0$.
Equation~(\ref{eqn:atwo}) defines $\vec{M}$ apart from a gradient field
and an arbitrary constant,
which are both chosen to be zero.
Consequently $\vec{M}$ vanishes outside the sample
and can therefore be interpreted 
as a magnetisation,
which is confined to the sample volume
and establishes the current density $\vec{J}$
by spatial variations.
Ampere's law results now in a homogeneous equation
for the field $\vec{\Omega}$

\begin{equation}
  \vec{\nabla}\times\vec{\Omega}=\vec{J}-\vec{\nabla}\times\vec{M}=0\,,
  \label{eqn:athree}
\end{equation}

which can thus be derived from a scalar potential

\begin{equation}
  -\vec{\nabla}\Phi=\vec{\Omega}\,.
  \label{eqn:afour}
\end{equation}

Taking the divergence of this expression 
and substituting the second Maxwell equation
($\vec{\nabla}\cdot\vec{B}=\vec{\nabla}\cdot(\vec{\Omega}+\vec{M})=0$)
results in

\begin{equation}
  \vec{\nabla}^2\Phi=\vec{\nabla}\cdot\vec{M}
  \label{eqn:afive}
\end{equation}

and the problem can be solved by the method of Green's functions

\begin{equation}
\Phi(\vec{r}\,)=
  -\frac{1}{4\pi}\int dV' 
  \frac{\vec{\nabla}'\cdot\vec{M}(\vec{r}\,')}{|\vec{r}-\vec{r}\,'|}
  +\frac{1}{4\pi}\oint d\vec{f}\,'\cdot
  \frac{\vec{M}(\vec{r}\,')}{|\vec{r}-\vec{r}\,'|}\,.
  \label{eqn:asix}
\end{equation}

The planar $z$-independent current flow
can be represented by $\vec{M}=M(x,y)\,\vec{e}_z$
and consequently the first term vanishes since
$\vec{\nabla}\cdot\vec{M}=\partial_zM_z=0$
in the sample volume.
Further,
only the surfaces perpendicular to $\vec{M}$
contribute to the second term. 
Using (\ref{eqn:asix}) the measured induction $B$
is expressed solely as a function of $M$ 

\begin{equation}
  B(\vec{r}\,) = \mu_0M(\vec{r}\,)
  +\underbrace{\frac{\mu_0}{4\pi}\int\!\!\int dx'dy'\,
  \frac{M(x',y')\,\Delta z}
       {(\Delta x^2+\Delta y^2+\Delta z^2)^{3/2}}\;
       \Big|^{d}_d+c}_{\Omega(\vec{r}\,)}.
  \label{eqn:aseven}
\end{equation}

Here $\Delta x=x-x',\ \Delta y=y-y',\ \Delta z=z-z'$ 
and the notation for evaluating antiderivatives are used
to indicate the positive contribution from the top ($z'=d$)
and the negative from the bottom ($z'=d+c$) surface.

Note,
that $\vec{\Omega}$ vanishes
for the case of an infinitely long slab (zero demagnetisation)
as $\Delta z\rightarrow \infty$ in~(\ref{eqn:aseven}).
In this case the induction is simply determined by
the magnitude of $\vec{M}$ at a certain position
and $\vec{B}=\mu_0\vec{M}$ inside and $\vec{B}=0$ outside the slab,
where $\vec{M}=0$.
However,
for finite geometries there will always be a contribution
from $\vec{\Omega}$ in (\ref{eqn:aseven})
and the relation between $\vec{B}$ and $\vec{M}$ is non-local.
Therefore the integral equation (\ref{eqn:aseven})
must be solved in order to attain $\vec{M}$.

The corresponding set of linear equations for the discrete measured data
is formulated by approximating the sample as an array of cubes with constant $M$.
Summing over all elements results in the matrix equation

\begin{equation}
  \sum_{k=1}^{n_x}\sum_{l=1}^{n_y} K_{i,j,k,l}M_{k,l}=\mu_0\Omega_{i,j}=B_{i,j}\,.
  \label{eqn:aeight}
\end{equation}

Here,
the matrix entries $K_{i,j,k,l}$ are calculated by
evaluating the antiderivative 

\begin{eqnarray}
   F(\Delta x,\Delta y,\Delta z) & = &
   \frac{\mu_0}{4\pi}\int\!\!\int dx\,dy\,
   \frac{\Delta z}{(\Delta x^2+\Delta y^2+\Delta z^2)^{3/2}}\\
   & = &\frac{\mu_0}{4\pi}\tan^{-1}(\frac{\Delta x\Delta y}
	{\Delta z\sqrt{\Delta x^2+\Delta y^2+\Delta z^2}})
   \label{eqn:anine}
\end{eqnarray}

at the eight corners of the cubes

\begin{equation}
  K_{i,j,k,l}  =  F(\Delta x,\Delta y,\Delta z)\; 
  \Big|^{s(i-k+\frac{1}{2})}_{s(i-k-\frac{1}{2})}\; 
  \Big|^{s(j-l+\frac{1}{2})}_{s(j-l-\frac{1}{2})}\;
  \Big|^{d+c}_{d}.
  \label{eqn:aten}
\end{equation}

Once the components $M_{k,l}$ are obtained by inverting~(\ref{eqn:aeight}),
the current density can be calculated by employing~(\ref{eqn:atwo}).
Further,
the induction at any distance $d$ outside the sample
can be obtained from~(\ref{eqn:aeight}).
Note,
that the magnetic induction $B$ in the central plane ($d=-c/2$) of the sample,
which is needed to obtain $J_c(B)$,
is given by

\begin{equation}
   \mu_0M_{i,j}+\sum_{k=1}^{n_x}\sum_{l=1}^{n_y} K_{i,j,k,l}M_{k,l}
   =\mu_0(M_{i,j}+\Omega_{i,j})=B_{i,j}
   \label{eqn:aeleven}
\end{equation}

as $M$ does not vanish inside the sample.

\end{document}